\documentclass{svjour2}                    
\smartqed  
\usepackage{graphicx}
 \journalname{J Low Temp Phys}
\begin{document}

\title{Exotic self-trapped states of an electron in superfluid helium
}

\titlerunning{Exotic electrons in helium}        

\author{Veit Elser}


\institute{ \at
              Department of Physics, Cornell University \\
              Tel.: +1-607-255-2340\\
              \email{ve10@cornell.edu}           
          }

\date{Received: date / Accepted: date}

\maketitle

\begin{abstract}
We explore the possibility that the fast and exotic negative ions in superfluid helium are electrons bound to quantized vortex structures, the simplest being a ring. In the states we consider, the electron energy is only slightly below the conduction band minimum of bulk helium. To support our proposal we present two calculations. In the first, we show that the electron pressure on the vortex core is insufficient to cavitate the helium and form an electron bubble. In the second, we estimate the equilibrium radius of the vortex ring that would bind an electron and find it is much smaller than the electron bubble, about 7 {\AA}. The many exotic ions reported in experiments might be bound states of an electron with more complex vortex structures.

\end{abstract}

\section{Introduction}
\label{intro}

Recent experiments by Maris and coworkers \cite{Wei2015} have strongly substantiated claims of ``fast" and ``exotic" negative ions in superfluid helium made over 40 years ago \cite{Doake1969,Ihas1971}. Not only do these latest experiments exactly reproduce the relative mobilities of some 14 species first identified by Ihas and Sanders \cite{Ihas1972}, they have resolved another four peaks in time-of-flight experiments as well as a signal arising from a continuously distributed mobility. That the same phenomena are observed with improved experimental techniques, including the use of carbon nanotube electrodes, argues strongly that these ions are intrinsic to the electron-helium system.

The elusive nature of these negative ions, whose mobility is up to a factor of six greater than that of the electron bubble, is compounded by the absence of a plausible model for their structure. The puzzle may be summarized as follows. Higher mobility implies smaller size, and suggests some entity that localizes a significant fraction of the electron wave function in a small volume. But the number of chemical species to be formed from helium atoms that could serve in this capacity are few and cannot account for the abundance of mobility peaks observed. A more speculative proposal, Maris' fissioned electron bubble \cite{Maris2000}, is not a tenable model as it is based on the adiabatic principle of quantum mechanics in a situation where it does not apply \cite{Elser2001,Mateo2010}.

This paper explores a proposal for the structures of the fast and exotic ions that appears to have been overlooked. Consider how these ions are created in the laboratory. Electrons are injected into liquid helium after they have diffused through the vapor above the liquid surface. Based on the energy parameter of the Meyer-Reif distribution \cite{Meyer1960} of the diffusing electron in a field, Wei \textit{et al.} estimate the electron energy in the range 3-6 eV at the time of injection \cite{Wei2015}. In the liquid the electron creates excitations, slowly losing energy and increasing its wavelength. When the energy has dropped to a particular value, $V_0\approx 1\,\mathrm{eV}$, the electron stops diffusing because below this energy the electron has no propagating states. This cutoff energy is usually referred to as the conduction band minimum of liquid helium.

In the adiabatic approximation, where the electron motion is so fast that helium atoms may be treated as static (in a configuration typical of the liquid), the constant $V_0$ corresponds to the mobility edge of Anderson localization. This interpretation of $V_0$ is better suited to our discussion, because the nature of the localized state accessed by the electron, even while the helium atoms are effectively static, determines the eventual structure of the ion.

The simplest geometry of the localized electron state is a roughly spherical wave function nestled in one of the larger cavities of the fluid. The normal electron bubble is formed, a self-trapped state, when on longer time scales the walls of this cavity are expanded by the zero-point pressure of the spherical electron wave function. Our proposal for the fast and exotic ions starts with the observation that the development of the localized state might be more complex and have alternate end-products.

Because the electron is much lighter than a helium atom, energy loss is a slow process and passage through the mobility edge is not abrupt. Hydrodynamical evolution of the fluid therefore does not switch on at a particular time, but develops by degree as the character of the electron wave function becomes increasingly localized. The hydrodynamical forcing by the electron can impart impulse to the fluid, a condition the superfluid can support only by being threaded with vorticity. Considering the fact that the energy of a 10 {\AA} radius vortex ring is only about 0.01 eV, the production of vorticity can be significant even for electrons that have lost just a fraction of the mobility edge energy of $V_0$.

The helium density at a vortex core is diminished and provides a line of weak attraction to the electron. And since in two dimensions an arbitrarily weak attractive potential binds a quantum particle, a straight vortex will confine the electron in its transverse plane. It follows that a localized distribution of vorticity of sufficient size, say a ring, will form a bound state with the electron. The electron energy for these states is only slightly below  $V_0$ because there is no large cavity that holds most of the electron wave function. We speculate that the fast and exotic negative ions may be explained by this class of bound state.

When modeling the final equilibrium state of the electron in the presence of vortices we encounter the following conundrum. In the Anderson localization picture, the electron sees the helium atoms as quenched disorder and has decaying wave function
\begin{equation}
\Psi\propto e^{-|\mathbf{r}|/\xi}
\end{equation}
where the correlation length
\begin{equation}
\xi\propto (V_0-E)^{-\nu}
\end{equation}
has exponent $\nu\approx 1$ \cite{Bauer1990}. This behavior, of the envelope of the wave function, applies to all the different realizations of ``disordered" helium atoms that arise in the fluid. On the other hand, in the conduction band picture used extensively in the past, bulk helium is treated as a continuum medium that imposes a uniform potential $V_0$ on the wave function. The decay of the wave function will then have exponent $\nu=1/2$. We have not tried to resolve this discrepancy and will adhere to the potential function model.

The best available modeling for the electron-helium interaction is based on density functionals \cite{Mateo2010}. This paper uses more elementary modeling and serves to motivate such a study. We carry out two calculations. In the first we show that an electron weakly bound to a linear vortex will not cavitate the core to transform it to a normal electron bubble. In the second, we find the equilibrium size of a vortex ring that binds an electron. Although the crudeness of our modeling of the vortex core introduces uncertainties, the resulting size is small and makes the ring a compelling candidate for the fast ion.

More complex vortex structures -- links and knots -- look like they might explain the multiplicity of ions observed in experiment. No calculations have been made in this paper to support that hypothesis. A class of structures that holds promise \cite{Maggioni2010} is the different ways a ring with multiple quanta of circulation may resolve itself into multiple filaments, possibly knotted and linked, all singly quantized. Owing to their parallel-oriented vorticity, such vortex tangles are less likely to reconnect and might support stable periodic motion. If the exotic ions corresponded to dynamic equilibrium structures such as this, they would be examples of true perpetual motion.

We have not explored extensions of the vortex-bound electron states that could explain the recently observed \cite{Wei2015} continuous component of the mobility distribution. However, the fissioning wave function proposal \cite{Wei2015} suffers from the same adiabaticity-breakdown concerns as its original application to photo-excited electron bubbles \cite{Maris2000}. In brief, Maris and coworkers consider the scenario where an electron entering the helium with energy exceeding $V_0$ becomes a superposition of reflected and transmitted waves, with amplitudes determined as in the textbook problem of a quantum particle encountering a fixed potential step. The transmitted part of the wave function then cavitates the helium to produce a bubble of variable size, consistent with the transmission amplitude. This analysis fails to acknowledge the fact that, in general, the electron-helium system must be described by a joint wave function. When the electron coordinate in this joint wave function is above the helium surface, the helium atoms are not compelled to preserve the bubble geometry that prevails when the electron coordinate is under the surface. The correct superposition is thus always a combination of normal electron bubble, and, a reflected-wave electron and cavity-free helium.

\section{Calculational framework and definition of scales}
\label{sec:1}
We model the helium as a continuum fluid of uniform mass density $\rho$ that forms a sharp interface at the empty cores of vortices. The interfacial surface energy, of a vortex ring of radius $R$ and core radius $a$, is
\begin{equation}
H_s=4\pi^2 R a\sigma,
\end{equation}
where $\sigma$ is the surface tension and we have assumed $R\gg a$. The energy of the helium in this model is just the kinetic energy in the corresponding classical flow field for vortex circulation $h/M$ ($M$ is the helium mass):
\begin{equation}\label{flow}
H_h=\frac{1}{2}\rho \left(\frac{h}{M}\right)^2 R\left(\log(8R/a)-2\right).
\end{equation}

The electron Hamiltonian is
\begin{equation}
H_e=-\frac{\hbar^2}{2 m}\nabla^2+V(\mathbf{r}),
\end{equation}
where $m$ is the electron mass and the potential $V$ is zero at points within distance $a$ of the vortex core and $V_0$ otherwise. We will let $V_0$ be our unit of energy and define our unit of length $a_0$ in terms of the associated kinetic energy:
\begin{equation}\label{a0}
V_0=\frac{\hbar^2}{2 m a_0^2}.
\end{equation}
For $V_0=1\, \mbox{eV}$, the scale $a_0=1.95$ {\AA} is comparable to the radius of a vortex core. Scaling the surface and flow energies in terms of our energy and length scales, we obtain the following two dimensionless parameters:
\begin{eqnarray}
\alpha&=&4\pi^2\left(\frac{\sigma a_0^2}{V_0}\right)=0.00326\\
\beta&=&\frac{1}{2}\rho\left(\frac{h}{M}\right)^2\left(\frac{a_0}{V_0}\right)=0.000878.\label{beta}
\end{eqnarray}

With our choice of energy and length scales the total Hamiltonian of the electron-vortex-ring system is
\begin{eqnarray}
H&=&H_e+H_s+H_h\\
&=&-\nabla^2+v(\mathbf{r})+\alpha R a+\beta R\left(\log(8R/a)-2\right),
\end{eqnarray}
where $v(\mathbf{r})=1$ except inside the tube of radius $a$ surrounding the ring, where it is zero. Although the magnitudes of the surface and flow terms are small due to their overall scales, we will see in the next Section that the ground state energy of $H_e$ is comparably small for the case of interest.

\section{Stability of the vortex core}
\label{sec:2}
In cylindrical coordinates appropriate to the ring geometry, the electron Hamiltonian has the form
\begin{equation}
H_e=H_0+H_1,
\end{equation}
where
\begin{eqnarray}
H_0&=&-\left(\frac{\partial^2}{\partial r^2}+\frac{\partial^2}{\partial z^2}\right)+v(r,z)\\
H_1&=&-\frac{1}{r}\frac{\partial}{\partial r}.\label{H1}
\end{eqnarray}
In this Section our interest is the core stability of a large ring where we can neglect the term $H_1$ which accounts for the curvature of the ring. This term is restored in the next Section where we determine the equilibrium radius of the ring.

In terms of the distance
\begin{equation}
s=\sqrt{(r-R)^2+z^2}
\end{equation}
from the (effectively straight) vortex core, the ground state electron wave function is
\begin{equation}
\Psi(s)\propto\left\{
\begin{array}{cc}
J_0(k s),& s<a\\
K_0(\kappa s),& s>a,\\
\end{array}
\right.
\end{equation}
with eigenvalue
\begin{equation}
H_0=k^2=1-\kappa^2.
\end{equation}
The value of $k$ is determined by matching the logarithmic derivative at $s=a$. Defining the variables $x=k a$ and $y=\kappa a$, the system of equations to be solved is
\begin{eqnarray}
x\frac{J_1(x)}{J_0(x)}&=&y\frac{K_1(y)}{K_0(y)}\label{bessel1}\\
x^2+y^2&=&a^2,
\end{eqnarray}
whose solutions are functions of the single parameter $a$. The electron energy is given by
\begin{equation}
H_0=\frac{x(a)^2}{a^2}.
\end{equation}

In the limit of small $a$, that is, cores smaller than the scale $a_0\approx 2$ {\AA}, equation (\ref{bessel1}) can be approximated as
\begin{equation}\label{matching}
\frac{x^2}{2}\approx \frac{-1}{\log{\left(\frac{e^\gamma}{2} y\right)}},
\end{equation}
where $\gamma$ is Euler's constant. Solving the system of equations in this approximation we find
\begin{equation}
\frac{x(a)^2}{a^2}\approx 1-(4/a^2) e^{-4/a^2-2\gamma},
\end{equation}
showing that the energy of the bound state (below the conduction band minimum) is exponentially small in the quantity $1/a^2$.

In Figure 1 we show the behavior of the total energy $H-V_0$ as a function of the core radius $a$ for two values of  $R$. The energy has a local maximum for $a$ of order 1 {\AA}, and decreases weakly for small $a$ and strongly for large $a$. For large $a$ the vortex core is expanded by the electron pressure, the fluid undergoes cavitation, and an electron bubble is formed. We are interested in the case of small $a$, where the electron pressure is exponentially feeble and the equilibrium structure is determined by the energetics of just the helium (balance of surface tension and Bernoulli pressures). If we ignore the electronic contribution entirely, the equilibrium core radius is given by
\begin{equation}
a=\beta/\alpha=0.269,
\end{equation}
or $0.53 \mbox{\AA}$ in physical units. An electron that finds itself in the vicinity of a vortex that is already small could not use it as a cavitation nucleus. The thermal activation barrier, for cavitation by uniform expansion of the core, is about 10 K for a 5 {\AA} ring and about five times this value for a 20 {\AA} ring. 
Although it has negligible effect on rings with small cores, an electron would nevertheless be bound when the ring is large enough. The effect of the bound electron on the size of the vortex ring is the task we take up in the next Section. In these calculations we treat the core radius as a fixed phenomenological parameter.

\begin{figure*}
\includegraphics[width=1.\textwidth]{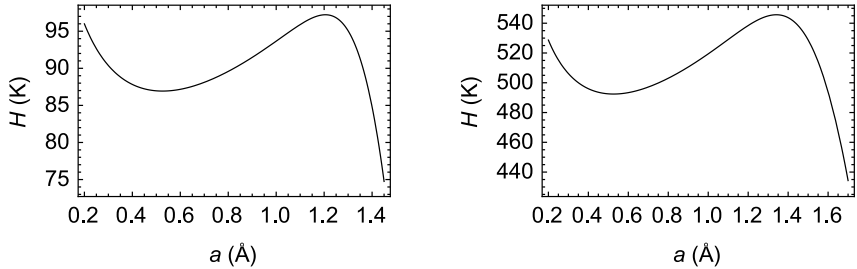}
\caption{Excess energy above the conduction band minimum of the electron-vortex-ring bound state as a function of core radius for two sizes of ring: $R=5$ {\AA} (left) and $R=20$ {\AA} (right).}
\label{fig:1}
\end{figure*}

\section{Electron-vortex-ring equilibrium}
\label{sec:3}

In the previous Section we saw that binding of an electron to a vortex lowers its energy, below the conduction band minimum, by an amount that behaves exponentially with the core radius. This was in the limit of a large ring, where the transverse extent of the electron wave function is small compared with the ring's radius of curvature. Since the electron wave function falls off exponentially, the effects of ring curvature on the binding energy is exponential as well. Finding the lower limit on the ring size, below which it no longer can bind an electron, is therefore a delicate mathematical exercise.

Two corrections must be applied to the calculation of Section~\ref{sec:2}. The first is to recognize that for the quasi-2D Hamiltonian $H_0$, an image vortex-core must be included at $r=-R$ in the $r$-$z$ plane in order to satisfy the boundary condition
\begin{equation}
\left.\frac{\partial\Psi}{\partial r}\right|_{r=0}=0.
\end{equation}
Writing the center of the true vortex (in the $r$-$z$ plane) as $\mathbf{R}_1$ and the image as $\mathbf{R}_2$, the electron wave function outside the core is
\begin{equation}\label{psiout}
\Psi_\mathrm{out}(\mathbf{r})=A\left(K_0(\kappa|\mathbf{r}-\mathbf{R}_1|)+K_0(\kappa|\mathbf{r}-\mathbf{R}_2|)\right),
\end{equation}
where $\mathbf{r}$ is a point in the $r$-$z$ plane and $A$ is the normalization constant. Since most of the wave function is outside the core in the case of interest, we determine $A$ by ignoring the contribution from the interior wave function. It will prove to be very convenient to express all the calculations in this Section in terms of the dimensionless parameter
\begin{equation}
q=\kappa R.
\end{equation} 
The normalization condition takes the form
\begin{equation}
1=A^2 R^3 f(q),
\end{equation}
where
\begin{equation}
f(q)=4\pi\int_0^\infty dy\int_0^\infty x dx \left(K_0(q\sqrt{(x-1)^2+y^2})+K_0(q\sqrt{(x+1)^2+y^2})\right)^2.
\end{equation}

The second correction in the binding energy calculation is the term $H_1$ of the electron Hamiltonian (\ref{H1}).
Treating this term as a first-order perturbation to be applied to the unperturbed wave function (\ref{psiout}), we obtain the correction
\begin{eqnarray}
\langle H_1\rangle &=& -2\pi\int_{-\infty}^\infty dz \int_0^\infty r dr \left(\Psi \frac{1}{r} \frac{\partial \Psi}{\partial r}\right)\\
&=&\pi\int_{-\infty}^\infty dz\;\Psi_\mathrm{out}^2(r=0,z)\\
&=&\frac{g(q)}{f(q)}\frac{1}{R^2},
\end{eqnarray}
where
\begin{equation}
g(q)=8\pi \int_0^\infty dy\; K_0^2(q\sqrt{1+y^2}).
\end{equation}

Combining the first two orders in the computation of the electron energy, we obtain
\begin{equation}
H_e=H_0+\langle H_1\rangle=1+\left(-q^2+\frac{g(q)}{f(q)}\right)\frac{1}{R^2(q)}.
\end{equation}
We will see shortly how the ring radius $R$ is also a function of $q$, owing to wave function matching at the core radius. The sign of the multiplier of $1/R^2$ determines whether the electron is bound. Numerically we find there is binding for $q>0.38$. However, the true bound on $q$ is smaller because we know the second order correction of a ground state energy (not calculated here) is always negative.

The wave function matching condition, in the case of a small core, is the same as equation (\ref{matching}) but with an extra term arising from the image core:
\begin{equation}
\frac{(ka)^2}{2}=\frac{-1}{\log{\left(\frac{e^\gamma}{2}\kappa a\right)}-K_0(2\kappa R)}.
\end{equation}
Using $\kappa R=q$ and $\kappa a=q (a/R)$, and solving for $R$ we obtain
\begin{equation}
R=\frac{e^\gamma}{2} q\, e^{-K_0(2q)}\;a e^{2/(ka)^2}.
\end{equation}
The electron wave vector in the core, $k$, depends weakly on $R$ through the relation
\begin{equation}
k^2=1-\kappa^2=1-(q/R)^2,
\end{equation}
from which we get the expansion
\begin{equation}
e^{2/(ka)^2}=e^{2/a^2}\left(1+\frac{2q^2}{(R a)^2}+\cdots\right).
\end{equation}
For the range of ring sizes and the parameter $q$ we will need, the correction terms are negligible and we can use the approximation
\begin{eqnarray}
R(q)&\approx& \frac{e^\gamma}{2} q\, e^{-K_0(2q)}\;a e^{2/a^2}\\
&=&\frac{1}{\sqrt{\delta}}\;q\, e^{-K_0(2q)},\label{radius}
\end{eqnarray}
where we have introduced a small parameter associated with the electron energy:
\begin{equation}\label{delta}
\delta=\frac{4}{a^2}e^{-4/a^2-2\gamma}.
\end{equation}
The electron binding energy is now expressed in terms of this scale and a function of $q$:
\begin{equation}\label{binding}
H_e-1=\delta \left(-q^2+\frac{g(q)}{f(q)}\right)\frac{e^{2 K_0(2q)}}{q^2}.
\end{equation}

The possibility of the electron stabilizing a vortex ring against shrinking down to zero radius depends critically on the parameter $\delta$. By (\ref{delta}) the latter depends sensitively on the dimensionless core radius $a$, and to a lesser extent on the conduction band minimum $V_0$ through the definition of the length scale $a_0$ (\ref{a0}). In the most optimistic scenario, when both of these are large, $\delta$ is largest and is best able to compete with the flow and surface tension energies of the vortex. For example, if we take $a=1.5$ {\AA} and $V_0=1.3$ eV, then $\delta=0.009$. This number is reduced by about one order of magnitude if the core radius is decreased to $1.2$ {\AA}.

\begin{figure}
\includegraphics[width=1.\textwidth]{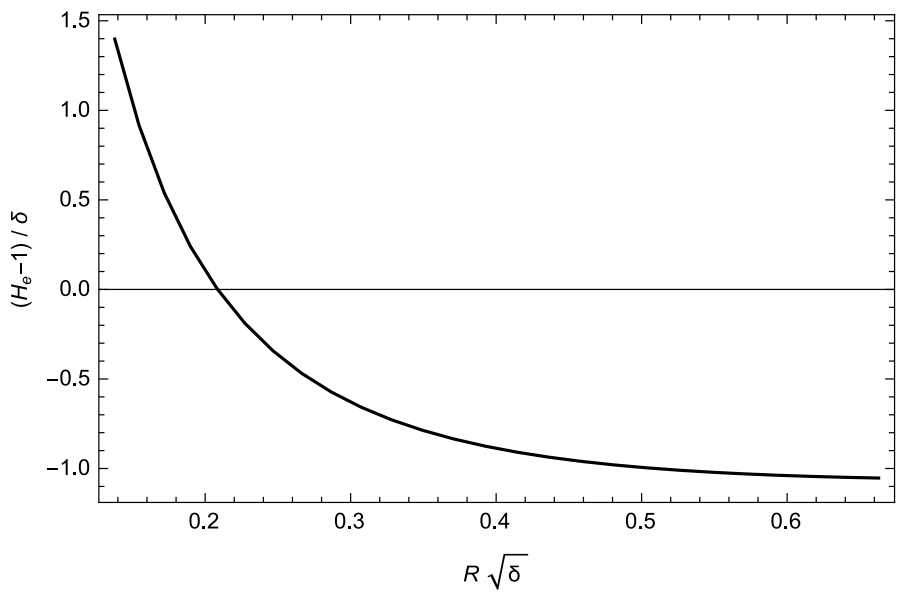}
\caption{Scaled electron binding energy (\ref{binding}) and corresponding scaled vortex ring radius (\ref{radius}) plotted as functions of the parameter $q$. The electron is bound to the ring for $q>0.38$ and $R>0.208/\sqrt{\delta}$.}
\label{fig:2}
\end{figure}

Figure 2 shows the electron binding energy (\ref{binding}) vs. ring radius (\ref{radius}) plotted parametrically as functions of $q$ and scaled by their respective powers of $\delta$. Although the details are more subtle than the  case of the normal bubble, the electron zero-point energy favors a large ring. It is this weak lowering of the energy with ring size that competes against the helium energy, which we take to be the energy of the flow in a ring vortex (\ref{flow}), where the core-surface energy contribution is included by adjusting the core radius $a$. The flow energy depends weakly on $a$ as well as $V_0$, through the scale $a_0$ (\ref{a0}) and the parameter $\beta$ (\ref{beta}). We choose $a=1$ {\AA} and $V_0=1$ eV, giving $\beta\approx 0.0009$. Figure~3 shows the total electron-vortex-ring energy (relative to $V_0$) vs. ring radius for $\delta=0.006$. The equilibrium value of the radius is $R=6.7$ {\AA}. In Table 1 we give binding energies, radii, and the parameter $q$ for equilibrium rings at other values of $\delta$.
These results show that when an equilibrium bound state exists, the vortex ring has radius $6.7$ {\AA}, and that probably no such state exists when $\delta$ is below $0.002$, because then $q$ is below the value for binding.

The electron energy and the hydrodynamic energy of the flowing helium are sensitive to very different characteristics of the vortex core. To model the former we introduced $\delta$, and treated it as an independent parameter even though it is nominally related to the hollow-core radius $a$ of the hydrodynamic model through equation (\ref{delta}). To make the best case for the electron-vortex bound state we selected optimal values for these parameters. Better modeling by density functional methods will eliminate this freedom and provide a more critical test of the proposed ion model.

\begin{figure}[t!]
\includegraphics[width=1.\textwidth]{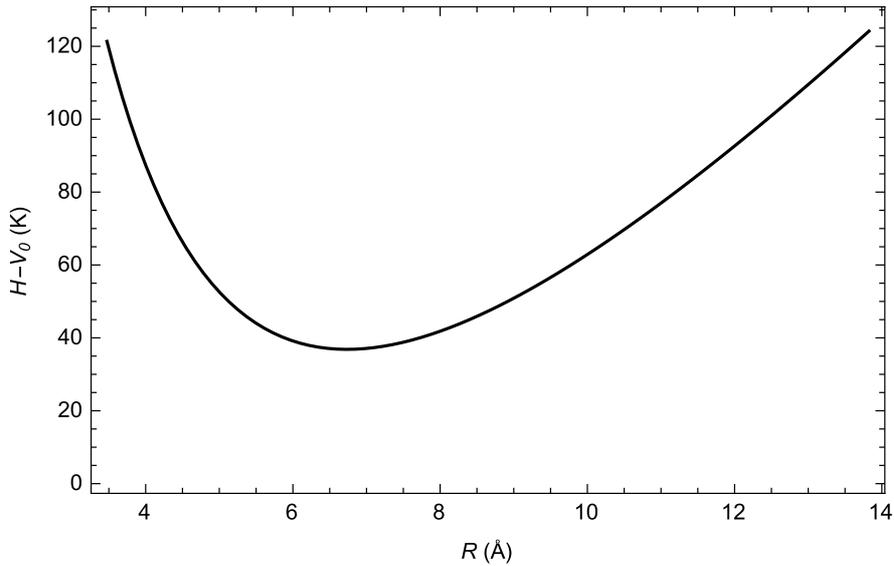}
\caption{Total energy of the electron-vortex-ring bound state as a function of ring radius for the case $\delta=0.006$. Equilibrium radii for various $\delta$ are given in Table 1.}
\label{fig:3}
\end{figure}

\begin{table}[b!]
\caption{Properties of equilibrium electron-vortex-ring bound states for various values of the parameter $\delta$.}
\label{tab:1}       
\begin{tabular}{c|ccc}
\hline\noalign{\smallskip}
$\delta$ & $H-V_0$ (K) & $R$ (\AA) & $q$  \\
\noalign{\smallskip}\hline\noalign{\smallskip}
0.010 & -20.2 & 6.72 & 0.515 \\
0.008 & 8.3 & 6.74 & 0.482 \\
0.006 & 36.8 & 6.73 & 0.440 \\
0.004 & 65.0 & 6.66 & 0.388 \\
0.002 & 90.8 & 6.38 & 0.310 \\
\noalign{\smallskip}\hline
\end{tabular}
\end{table}

\section{Creation and transport of electron-vortex-ring bound states}

There can be no doubt that an electron injected into liquid helium above the conduction band minimum can be captured as a bound state to a large enough vortex ring (or something topologically more complex). As long as the vortex core radius is below some minimum, of the same order as a free vortex, the electron pressure is insufficient to cavitate the core and form an electron bubble. What happens after capture is less certain.

Because the total energy of the electron-vortex-ring bound state (calculated in Section \ref{sec:3}) is an increasing function of radius for large radius, over time the ring will decrease its radius through emission and scattering of phonons and rotons. The most intriguing scenario is that this process will cease, with equilibrium  established when the ring has radius of order $7$~{\AA}. This possibility rests on the reliability of our bound state energy estimate, which unfortunately scales with the small parameter $\delta$ whose value is sensitively dependent on the core radius and the conduction band minimum, $V_0$. When $\delta$ falls below about 0.002, the electron becomes unbound ($H_e>V_0$) and the ring is free to decay.

The scenario of no equilibrium electron-vortex-ring state has trouble explaining the transport measurements of the fast and exotic ions. There are two cases to be considered. For large electric field $E$, when the electric force is balanced by a drag force proportional both to the radius $R$ and velocity $v$ of the ring, we get the relations
\begin{equation}
E\propto R v,
\end{equation}
\begin{equation}
v\propto \frac{1}{R}\log{R}\propto E e^{-E/E_0},
\end{equation}
where we have used the velocity $v$ of a ring advected by its own flow field. In the weak field case, when the drag must be modeled as stochastic, the ring can decay before its impulse and size are restored by the electric force. Now the electron is liberated and free to be recaptured by vorticity elsewhere in the helium. Thus, while a decaying-exponential drift velocity is consistent with the high field measurements of Eden and McClintock \cite{Eden1984}, the low field mobility would appear to lack the robust characteristics recently established by Maris and coworkers \cite{Wei2015}.

The most promising scenario then, of electron-vortex-bound states as the elusive fast and exotic ions, is for the electron to establish an equilibrium structure. These would be metastable states, as the electron energy is nearly 1 eV above the energy of the ground state electron bubble. Nevertheless, their stability is made possible by two things. First, because the electron is in the ground state with respect to the potential function of the vortex-threaded helium, the state cannot decay radiatively. Second, decay via tunneling to an expanded core that can cavitate is suppressed by the relatively large size of the equilibrium ring.

The starting point for discussing the transport properties of an equilibrium ring is to realize that such a ring would have zero velocity (in the absence of a field) even while its impulse $p$ is non-zero. By general principles,
\begin{equation}
v=\frac{\partial H}{\partial p}=\frac{\partial H}{\partial R}\frac{\partial R}{\partial p}=0,
\end{equation}
since the total energy is stationary at the equilibrium ring radius. The cartoon of flow fields in Figure 4 shows how this can be understood in terms of force balance. The two terms in $H$ contribute opposing generalized forces:
\begin{equation}
\frac{\partial H}{\partial R}=\frac{\partial H_h}{\partial R}+\frac{\partial H_e}{\partial R}=-F_h-F_e.
\end{equation}
The first force is the inward Magnus force per unit length of vortex times the length of the ring. This is balanced by the outward force produced by the bound electron.

Like the ordinary electron bubble, the zero-velocity electron-ring bound state acquires a finite drift velocity in the presence of a field. Two features suggest the drag on the ring, due to phonon-roton scattering, would be significantly smaller for the ring than the electron bubble: (i) its size (radius) is smaller, (ii) the ``helium contrast" is much smaller (empty core vs. empty bubble). On the other hand, unlike the electron bubble, the vortex ring is surrounded by a non-zero flow field even in the limit of zero velocity. This has the effect of enhancing the effective scattering cross section.

\begin{figure}[t!]
\includegraphics[width=1.\textwidth]{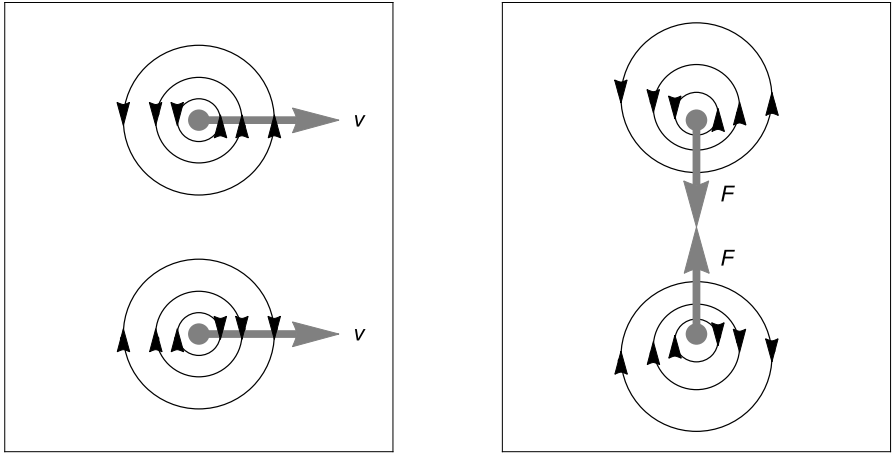}
\caption{Contrasting flow fields around a free vortex ring (left), and a ring that binds an electron (right). The free ring is advected by its own flow field and moves to the right with velocity $v$ as shown. Because the core is at rest with respect to the flow, the flow around it is uniform and there are no Bernoulli forces. The flow pattern around the $v=0$ ring that binds an electron, on the right, is obtained by transforming the free ring on the left to the frame moving with velocity $v$. This results in a flow asymmetry around the core and unequal Bernoulli pressures that produce a net inward Magnus force $F$. Equilibrium of the core is maintained because the bound electron generates a net outward force of equal magnitude.}
\label{fig:4}
\end{figure}

\section{Conclusion}

It is known that vortices can bind electron bubbles. The bound states considered in this paper are qualitatively different in that the electron leaves a much smaller footprint in the helium. As a possible model of the fast negative ion, we studied the binding of an electron to a vortex ring, where the effect of the electron is simply to stabilize the ring against decay when its radius is about 7 {\AA}. The helium density in this proposed ion is much more uniform than when the electron forms a bubble. Although the vortex ring's flow field extends the range over which it scatters phonons and rotons, the small size and more uniform density of the proposed ion would increase its mobility significantly over that of the electron bubble.

The most perplexing experimental fact that any model of the exotic ions must address is their great number. Although we have analyzed in detail only the vortex ring, the same stabilization mechanism should apply to topologically more complex vortex structures, such as knots and links. As in the case of the ring, the electron's energy would be only slightly below the conduction band minimum. Even so, the energetics of the electron state is such that it favors straight over curved vortex filaments, the net effect being to expand the vortex ``tangle". The density functional methodology \cite{Mateo2010}, suitably adapted, might be able to test the validity of this mechanism and if confirmed, identify the most stable structures.

\begin{acknowledgements}
I thank A. Fetter and H. Maris for discussions and especially S. Dutta for his critical reading of the manuscript.
\end{acknowledgements}



\end{document}